\documentclass[man,12pt,letterpaper]{apa7}
\usepackage[T1]{fontenc}
\usepackage{lmodern}
\usepackage{microtype}
\usepackage{ragged2e}
\usepackage[american]{babel}
\usepackage{csquotes}
\usepackage[style=apa,backend=biber]{biblatex}
\usepackage{hyperref}
\hypersetup{hidelinks,bookmarksnumbered}
\makeatletter
\let\apasection\section
\renewcommand{\section}{%
  \@ifstar{\apasection*}{%
    \@startsection{section}{1}{\z@}%
      {\b@level@one@skip}{\e@level@one@skip}%
      {\normalfont\normalsize\bfseries\raggedright}}}
\renewcommand{\subsection}{%
  \@startsection{subsection}{2}{\z@}%
    {\b@level@two@skip}{\e@level@two@skip}%
    {\normalfont\normalsize\bfseries\raggedright}}
\newcommand{\section@headingnumber}{\thesection.}
\renewcommand{\@seccntformat}[1]{%
  \@ifundefined{#1@headingnumber}{\csname the#1\endcsname}{\csname #1@headingnumber\endcsname}%
  \enspace}
\makeatother

\usepackage{amsmath,amssymb,amsthm,mathtools}
\usepackage{needspace}

\DeclareMathOperator{\Cov}{Cov}
\DeclareMathOperator{\Cor}{Cor}
\DeclareMathOperator{\tr}{tr}
\DeclareMathOperator{\vecop}{vec}
\DeclareMathOperator{\vech}{vech}
\DeclareMathOperator{\rank}{rank}
\newcommand{\dd}{\mathrm{d}}
\newcommand{\E}{\mathbb{E}}
\newcommand{\R}{\mathbb{R}}
\newcommand{\one}{\mathbf{1}}
\newcommand{\Dp}{\mathsf{D}_p}

\newtheorem{theorem}{Theorem}

\title{The Random Intercept Cross-Lagged Panel Model as a Statistical Model: Identification, Estimation, and Model Comparison}
\shorttitle{The RI-CLPM as a Statistical Model}
\authorsnames{Zihao Cao}
\authorsaffiliations{East China Normal University}

\authornote{
  \addORCIDlink{Zihao Cao}{0009-0002-0962-7849}

  Correspondence concerning this article should be addressed to Zihao Cao,
  East China Normal University. E-mail: \href{mailto:10180350247@stu.ecnu.edu.cn}{10180350247@stu.ecnu.edu.cn}

  The author used ChatGPT (OpenAI, GPT-6 Astra) for English-language editing and assistance in checking the mathematical derivations and code; the core ideas and derivations are the author’s own, and the author takes full responsibility for the content of this work.
}

\abstract{The random intercept cross-lagged panel model (RI-CLPM) distinguishes stable differences between people from lagged associations within individuals. We use a matrix representation to study its statistical properties for \(p\) variables measured across \(T\) waves.
We derive a necessary and sufficient rank condition for global identification when transition matrices and innovation covariances vary by wave. Under time-invariant transitions, the condition specifies when three waves suffice. For complete Gaussian trajectories, matrix differentiation yields the profile likelihood score for each covariance and transition block.
We also derive the population parameters of the conventional cross-lagged panel model (CLPM) when an RI-CLPM with fixed parameters generates the data. Our formulas quantify how CLPM slopes for observed variables differ from RI-CLPM slopes for within-person deviations at each transition. When the RI-CLPM transition matrix is time-invariant, the comparison also covers CLPM slopes pooled across waves. The formulas recover earlier bivariate results and show when cross-lagged slopes can appear, cancel, reverse sign, or change their ordering across models.}

\begin{document}
\maketitle

\setlength{\JustifyingParindent}{\parindent}
\justifying
\tolerance=1500
\setlength{\emergencystretch}{1.5em}
% Begin chapters/01_model.tex
\section{Introduction}
\label{sec:introduction}

Psychologists increasingly use longitudinal data to examine how psychological processes change over time \parencite{Hopwood2022ConnectingTheoryMethods,Nestler2023UnivariateARasMixed}. Such questions often require separating stable differences among people from changes within the same person. The random intercept cross-lagged panel model (RI-CLPM) extends the conventional cross-lagged panel model (CLPM) by using random intercepts to represent stable between-person variation \parencite{Hamaker2015RICLPM}. Its lagged slopes therefore describe associations between departures from each person's expected levels. When stable differences contribute to observed covariances, the two models can give different answers about which lagged associations are present, their signs, and their relative magnitudes \parencite{Hamaker2015RICLPM,Hamaker2026WithinBetween}.

Prior work shows how to extend the RI-CLPM for applied questions and how to plan sample sizes and wave counts \parencite{Mulder2021ThreeExtensions,Mulder2023powRICLPM}. An analysis of bullying and internalizing problems illustrates how accounting for within-wave associations can change cross-lagged estimates \parencite{Speyer2024ConcurrentRICLPM}. Relatively few studies have treated the RI-CLPM as a statistical model in its own right and derived its properties analytically.

First, we use the block-tridiagonal within-person precision matrix to derive a necessary and sufficient rank condition for global identification. The condition applies to \(p\) variables measured across \(T\) waves, with both transition matrices and innovation covariances free to vary across waves. The time-invariant case recovers the three-wave minimum described by \textcite{Hamaker2015RICLPM}. With at least three waves and a common transition matrix \(F\), global identification holds exactly when \(I_p-F\) is nonsingular. Second, for complete Gaussian trajectories, we profile out the wave means and derive exact blockwise maximum likelihood score equations. These equations provide a starting point for studying alternative estimators. Third, given an RI-CLPM with fixed parameters as the data-generating process, we use its covariance blocks to derive the population CLPM slopes and their differences from the within-person transition slopes. The results cover wave-specific and pooled slopes and recover earlier bivariate comparisons \parencite{Hamaker2015RICLPM,RobitzschLudtke2026}.

\section{The Random Intercept Cross-Lagged Panel Model}
\label{sec:model}

The model draws on the random-intercept decomposition of \textcite{Hamaker2015RICLPM} and the unified longitudinal modeling framework of \textcite{Usami2019UnifiedFramework}. We write it in matrix form for \(p\) variables measured at \(T\) waves, with transition matrices and innovation covariances allowed to vary by wave. The observation and latent-state equations determine the covariance of the stacked measurements and its individual wave blocks. We also specify the parameters and differential notation used in the derivations.

\subsection{Observations and Latent Components}

Write \(y_{it}\in\R^p\) for the observation vector for individual \(i\) at wave \(t\), with \(i=1,\ldots,N\), \(t=1,\ldots,T\), \(p\geq 2\), and \(T\geq 2\). Allow the wave mean \(\mu_t\in\R^p\) to vary freely. The model equations are
\begin{align}
y_{it}&=\mu_t+\eta_i+w_{it}, &&t=1,\ldots,T,
\label{eq:observation}\\
w_{it}&=F_t w_{i,t-1}+\varepsilon_{it}, &&t=2,\ldots,T.
\label{eq:within}
\end{align}
The time-invariant random intercept is \(\eta_i\in\R^p\), the within-person deviation is \(w_{it}\in\R^p\), and the transition matrix \(F_t\in\R^{p\times p}\) relates wave \(t-1\) to wave \(t\). Entry \((F_t)_{jk}\) is the lagged slope from within-person component \(k\) to component \(j\), conditional on the other components at the preceding wave. The diagonal entries of \(F_t\) are autoregressive slopes, and its off-diagonal entries are cross-lagged slopes. We use this row and column convention throughout.

For each individual \(i\), the random inputs \(\eta_i,w_{i1},\varepsilon_{i2},\ldots,\varepsilon_{iT}\) are mutually independent and follow Gaussian distributions:
\begin{equation}
 \begin{gathered}
   \eta_i\sim\mathcal{N}_p(0,\Omega),\qquad
   w_{i1}\sim\mathcal{N}_p(0,\Gamma),\qquad
   \varepsilon_{it}\sim\mathcal{N}_p(0,\Psi_t),\\
   \Omega,\Gamma,\Psi_t\succ 0.
 \end{gathered}
 \label{eq:inputs}
\end{equation}
Trajectories are independent and identically distributed across individuals. The covariance \(\Omega\) describes stable between-person differences, \(\Gamma\) describes within-person deviations at wave \(1\), and \(\Psi_t\) is the covariance of the innovations at wave \(t\).

The first-wave specification requires care when the process starts before observation. We assign \(w_{i1}\) a free covariance matrix \(\Gamma\) and assume independence from the random intercept and later innovations. \textcite{Bollen2026ALTAssessment} discuss alternative initial conditions for autoregressive latent trajectory (ALT) models.

Iterating \eqref{eq:within} writes each \(w_{it}\) as a fixed linear combination of \(w_{i1},\varepsilon_{i2},\ldots,\varepsilon_{it}\). It follows that \(\eta_i\) is independent of the full within-person trajectory \((w_{i1},\ldots,w_{iT})\), so the stable trait is orthogonal to the within-person state at every wave. The unrestricted means \(\mu_t\) account for systematic shifts in the population means across waves \parencite{Hamaker2020FixedRandom,Usami2022Interpretation}.

\subsection{Stacked Covariance and Block Structure}

Stacking the wave vectors in chronological order, let \(y_i=(y_{i1}^{\top},\ldots,y_{iT}^{\top})^\top\), and define \(\mu\) and \(w_i\) analogously. Set \(d=pT\) and define the stacked input vector \(v_i=(w_{i1}^\top,\varepsilon_{i2}^\top,\ldots,\varepsilon_{iT}^\top)^\top\). In matrix form, \(\mathcal{L} w_i=v_i\) and \(\mathcal{D}=\Cov(v_i,v_i)\), where
\begin{equation}
\mathcal{L}=
\begin{pmatrix}
 I_p&0&\cdots&0&0\\
 -F_2&I_p&\cdots&0&0\\
 0&-F_3&\ddots&\vdots&\vdots\\
 \vdots&\ddots&\ddots&I_p&0\\
 0&\cdots&0&-F_T&I_p
\end{pmatrix},
\qquad
\mathcal{D}=
\begin{pmatrix}
 \Gamma&0&\cdots&0\\
 0&\Psi_2&\cdots&0\\
 \vdots&\vdots&\ddots&\vdots\\
 0&0&\cdots&\Psi_T
\end{pmatrix}.
\label{eq:stacked-inputs}
\end{equation}
The block matrices extend over the \(T\) observed waves. The lower block-bidiagonal matrix \(\mathcal{L}\) has diagonal blocks equal to \(I_p\), so \(\mathcal{L}\) is invertible and the recursion has a unique solution. This gives
\begin{align}
 K&=\Cov(w_i,w_i)=\mathcal{L}^{-1}\mathcal{D}(\mathcal{L}^{-1})^\top,
 \label{eq:K}\\
 y_i&=\mu+(\one_T\otimes I_p)\eta_i+w_i,
 \qquad y_i\sim\mathcal{N}_d(\mu,\Sigma),\nonumber\\
 \Sigma&=K+(\one_T\one_T^\top)\otimes\Omega.
 \label{eq:Sigma}
\end{align}
By congruence in \eqref{eq:K}, \(K\succ 0\); since \((\one_T\one_T^\top)\otimes\Omega\succeq 0\), it follows that \(\Sigma\succ 0\) as well. Throughout, we denote the within-person precision matrix by \(Q=K^{-1}\) and the observed precision matrix by \(P=\Sigma^{-1}\).

Turning to individual \(p\times p\) blocks, define \(K_{ts}=\Cov(w_{it},w_{is})\) and \(\Sigma_{ts}=\Cov(y_{it},y_{is})\). Because every wave shares the same random intercept \(\eta_i\) and the wave means are deterministic, the observed covariance blocks decompose additively as
\begin{equation}
 \Sigma_{ts}=\Omega+K_{ts}.
 \label{eq:observed-block}
\end{equation}
At the initial wave, \(K_{11}=\Gamma\). For \(s<t\), taking the covariance of both sides of \eqref{eq:within} with \(w_{is}\) eliminates the innovation term \(\varepsilon_{it}\) due to orthogonality with past states. Similarly, taking the covariance of \eqref{eq:within} with itself eliminates the transition--innovation cross terms. These calculations yield the block recursions
\begin{align}
 K_{tt}&=F_tK_{t-1,t-1}F_t^\top+\Psi_t,\label{eq:K-diag}\\
 K_{ts}&=F_tK_{t-1,s}
       =F_tF_{t-1}\cdots F_{s+1}K_{ss},\quad t>s,
 \qquad K_{st}=K_{ts}^\top.
 \label{eq:K-lag}
\end{align}
The following adjacent-wave covariance blocks determine the population CLPM slope in Section~\ref{sec:comparison}
\begin{equation}
 \Sigma_{t-1,t-1}=\Omega+K_{t-1,t-1},
 \qquad
 \Sigma_{t,t-1}=\Omega+F_tK_{t-1,t-1}.
 \label{eq:adjacent-blocks}
\end{equation}

In the unrestricted model, \(F_t\) and \(\Psi_t\) vary freely across waves. Time-invariant transitions impose \(F_t=F\), and time-invariant innovations impose \(\Psi_t=\Psi\). Stationarity is a natural requirement when both quantities are constant. It requires \(\rho(F)<1\) and \(\Gamma=F\Gamma F^\top+\Psi\), where \(\rho(\cdot)\) is the spectral radius. The discrete Lyapunov equation gives \(K_{tt}=\Gamma\) for every wave by \eqref{eq:K-diag}. Treating constant paths and stationary initialization as separate restrictions makes the weighting across waves explicit, which matters for model specification and study design \parencite{Mulder2021ThreeExtensions,Mulder2023powRICLPM}.

\subsection{Parameters and Differential Convention}

Let \(s_p=p(p+1)/2\). We denote columnwise vectorization of general matrices by \(\vecop(\cdot)\) and lower-triangular half-vectorization of symmetric matrices by \(\vech(\cdot)\). The covariance parameter vector is collected as
\begin{equation}
 \theta=
 \begin{pmatrix}
 \vech(\Omega)\\
 \vech(\Gamma)\\
 \vech(\Psi_2)\\
 \vdots\\
 \vech(\Psi_T)\\
 \vecop(F_2)\\
 \vdots\\
 \vecop(F_T)
 \end{pmatrix}
 \in \R^{(T+1)s_p+(T-1)p^2},
 \label{eq:theta}
\end{equation}
while the \(d\) mean parameters in \(\mu\) are treated separately. The duplication matrix \(\Dp\in\R^{p^2\times s_p}\) satisfies \(\vecop(M)=\Dp\vech(M)\) for any symmetric \(p\times p\) matrix \(M\) and has full column rank \(s_p\).

We adopt the matrix differential calculus convention of \textcite{MagnusNeudecker2019}, where \(\dd(\cdot)\) denotes the first-order differential induced by a parameter perturbation. For a scalar-valued function \(f(\theta)\), its column score vector \(U\) is defined via \(\dd f=U^\top\dd\theta\). Vector and matrix representations are linked through the standard identities
\begin{align}
 \vecop(AMB)&=(B^\top\otimes A)\vecop(M),\label{eq:vec-product}\\
 \tr(A^\top B)&=\vecop(A)^\top\vecop(B),\qquad
 \dd\vecop(M)=\Dp\,\dd\vech(M)\quad(M=M^\top).
 \label{eq:trace-duplication}
\end{align}
\textcite{AbadirMagnus2005} provide proofs of these identities.
% End chapters/01_model.tex
% Begin chapters/02_identification.tex
\section{Identification}
\label{sec:identification}

In a Gaussian structural equation model, a parameter vector is globally identified if no other admissible parameter vector implies the same mean vector and covariance matrix \parencite{Hunter2025SEMIdentification}.
For the RI-CLPM, consider two admissible covariance parameter vectors satisfying
\begin{equation}
 \Sigma(\theta^{(1)})=\Sigma(\theta^{(2)}).
 \label{eq:equal-sigma}
\end{equation}
The question is whether \eqref{eq:equal-sigma} forces \(\theta^{(1)}=\theta^{(2)}\). The observed distribution already determines the wave means, so identification here concerns the covariance parameters.

The identity \(\Sigma_{ts}=\Omega+K_{ts}\) means that the observed covariance \(\Sigma\) and a candidate \(\Omega\) fix every block of \(K=\Sigma-(\one_T\one_T^\top)\otimes\Omega\). We first show that admissible within-person parameters correspond one-to-one to positive definite block-tridiagonal matrices \(Q\). This lets us check whether changing \(\Omega\) can preserve positive definiteness and the required zero blocks of \(Q\).

\subsection{The Within-Person Precision and Inverse Parameter Map}

Inverting the congruence in \eqref{eq:K} yields the within-person precision matrix
\begin{equation}
 Q=K^{-1}=\mathcal{L}^\top\mathcal{D}^{-1}\mathcal{L}.
 \label{eq:Q-factor}
\end{equation}
From \eqref{eq:stacked-inputs}, \(\mathcal{L}^\top\) is upper block bidiagonal and \(\mathcal{D}^{-1}\) is block diagonal:
\begin{equation}
 \mathcal{L}^\top=
 \begin{pmatrix}
 I_p&-F_2^\top&0&\cdots&0\\
 0&I_p&-F_3^\top&\ddots&\vdots\\
 \vdots&\ddots&\ddots&\ddots&0\\
 0&\cdots&0&I_p&-F_T^\top\\
 0&\cdots&\cdots&0&I_p
 \end{pmatrix},
 \quad
 \mathcal{D}^{-1}=
 \begin{pmatrix}
 \Gamma^{-1}&0&\cdots&0\\
 0&\Psi_2^{-1}&\cdots&0\\
 \vdots&\vdots&\ddots&\vdots\\
 0&0&\cdots&\Psi_T^{-1}
 \end{pmatrix}.
 \label{eq:Q-factors-explicit}
\end{equation}
The \((t,s)\) block of \(Q\) is obtained by summing across the block rows \(r=1,\ldots,T\):
\begin{equation}
 Q_{ts}=\sum_{r=1}^{T}
    \mathcal{L}_{rt}^{\top}(\mathcal{D}^{-1})_{rr}\mathcal{L}_{rs}.
 \label{eq:Q-block-multiplication}
\end{equation}
Block column \(j\) of \(\mathcal{L}\) contains nonzero blocks solely in block rows \(j\) and (for \(j<T\)) \(j+1\). Consequently, whenever \(t\ge s+2\), columns \(t\) and \(s\) share no nonzero block row, so \(Q_{ts}=0_{p\times p}\). For \(s=t-1\), the sole overlapping row is \(r=t\), yielding \(I_p^\top\Psi_t^{-1}(-F_t)\). For \(t=s\), overlapping contributions arise from rows \(t\) and \(t+1\) (with the latter absent at \(t=T\)). Thus, the lower triangle of \(Q\) is given explicitly by
\begin{align}
 Q_{11}&=\Gamma^{-1}+F_2^\top\Psi_2^{-1}F_2,\label{eq:Q-first}\\
 Q_{tt}&=\Psi_t^{-1}+F_{t+1}^\top\Psi_{t+1}^{-1}F_{t+1},
              &&2\le t<T,\label{eq:Q-interior}\\
 Q_{TT}&=\Psi_T^{-1},\qquad
 Q_{t,t-1}=-\Psi_t^{-1}F_t, &&2\le t\le T,\label{eq:Q-terminal}\\
 Q_{ts}&=0_{p\times p},&&t\ge s+2,
 \label{eq:Q-zeros}
\end{align}
and symmetry supplies the upper triangle (\(Q_{st}=Q_{ts}^\top\)). The block-tridiagonal structure of \(Q\) follows from the first-order Markov property. The covariance \(K\) can still have nonzero blocks at every lag because effects propagate recursively in \eqref{eq:K-lag}.

Conversely, given an admissible within-person covariance \(K=Q^{-1}\), its constituent parameters are uniquely recovered from its blocks. Specifically, \(\Gamma=K_{11}\). For \(2\le t\le T\), taking the covariance of \(w_{it}=F_t w_{i,t-1}+\varepsilon_{it}\) with \(w_{i,t-1}\) yields \(K_{t,t-1}=F_t K_{t-1,t-1}\) by orthogonality of \(\varepsilon_{it}\). Right-multiplying by \(K_{t-1,t-1}^{-1}\) recovers \(F_t\), and \eqref{eq:K-diag} then yields \(\Psi_t\):
\begin{equation}
 \begin{aligned}
 F_t    &=K_{t,t-1}K_{t-1,t-1}^{-1},\\
 \Psi_t &=K_{tt}-F_tK_{t-1,t-1}F_t^\top,\qquad 2\le t\le T.
 \end{aligned}
 \label{eq:within-recovery-from-K}
\end{equation}
For example, at wave \(2\), \(K_{21}\) and \(K_{11}\) determine \(F_2\), and subtracting the propagated variance \(F_2 K_{11} F_2^\top\) from \(K_{22}\) isolates \(\Psi_2\).

To verify that every positive definite block-tridiagonal matrix \(Q\) corresponds to a unique admissible parameter set, solve \eqref{eq:Q-first}--\eqref{eq:Q-terminal} backward from wave \(T\). Since wave \(T\) has no outgoing transition, \eqref{eq:Q-terminal} determines \(\Psi_T=Q_{TT}^{-1}\succ 0\) and \(F_T=-\Psi_T Q_{T,T-1}\). The corresponding contribution to \(Q_{T-1,T-1}\) is
\begin{equation}
 F_T^\top\Psi_T^{-1}F_T
 =Q_{T-1,T}Q_{TT}^{-1}Q_{T,T-1}.
 \label{eq:Q-schur-contribution}
\end{equation}
The Schur complement of \(Q_{TT}\) in \(Q\) has order \((T-1)p\). Since \(Q_{Ts}=0\) for \(s\le T-2\), the only change to the leading principal submatrix is the subtraction of \eqref{eq:Q-schur-contribution} from \(Q_{T-1,T-1}\). Since \(Q\succ 0\), its Schur complement is positive definite and retains the block-tridiagonal structure. Repeating the step recovers one transition \(F_t\) and one positive definite innovation covariance \(\Psi_t\) per wave. The last \(p\times p\) block gives \(\Gamma^{-1}\succ 0\). Substituting the recovered parameters into \eqref{eq:Q-factor} reconstructs \(Q\) exactly, completing the one-to-one correspondence.

\subsection{Exact Covariance-Preserving Perturbations}

Let \((K,\Omega)\) denote the baseline decomposition and write an alternative decomposition as \((\widetilde{K},\widetilde{\Omega})\) with \(\widetilde{\Omega}=\Omega+\Delta\) for a symmetric perturbation \(\Delta=\Delta^\top\). Equality of the observed covariances requires
\begin{equation*}
 K+(\one_T\one_T^\top)\otimes\Omega
 =\widetilde{K}+(\one_T\one_T^\top)\otimes(\Omega+\Delta),
\end{equation*}
which simplifies to
\begin{equation}
 \widetilde{K}=K-(\one_T\one_T^\top)\otimes\Delta.
 \label{eq:K-perturb}
\end{equation}
Every entry of \(\one_T\one_T^\top\) is \(1\). Thus adding \(\Delta\) to \(\Omega\) subtracts \(\Delta\) from each \(K_{ts}\). The perturbed parameters are admissible if \(\widetilde{\Omega}\succ 0\), \(\widetilde{K}\succ 0\), and \(\widetilde{Q}=\widetilde{K}^{-1}\) has the zero-block pattern in \eqref{eq:Q-zeros}.

Using the resolvent identity \(\widetilde{Q}-Q=\widetilde{Q}(K-\widetilde{K})Q\), we substitute \eqref{eq:K-perturb} to obtain
\begin{equation*}
 \widetilde{Q}-Q
 =\widetilde{Q}\bigl[(\one_T\one_T^\top)\otimes\Delta\bigr]Q.
\end{equation*}
Because every block of \((\one_T\one_T^\top)\otimes\Delta\) equals \(\Delta\), the \((t,s)\) block of \(\widetilde{Q}-Q\) factorizes cleanly:
\begin{align*}
 \widetilde{Q}_{ts}-Q_{ts}
 &=\sum_{a=1}^T\sum_{b=1}^T\widetilde{Q}_{ta}\Delta Q_{bs}\\
 &=\left(\sum_{a=1}^T\widetilde{Q}_{ta}\right)\Delta
   \left(\sum_{b=1}^T Q_{bs}\right).
\end{align*}
Define the block row sums \(Z_t\) and \(\widetilde{Z}_t\) of \(Q\) and \(\widetilde{Q}\), along with the grand block sum \(\Lambda\) of \(Q\):
\begin{equation}
 Z_t=\sum_{s=1}^T Q_{ts},\qquad
 \widetilde{Z}_t=\sum_{s=1}^T\widetilde{Q}_{ts},\qquad
 \Lambda=\sum_{s=1}^T Z_s^\top.
 \label{eq:Z}
\end{equation}
Symmetry of \(Q\) ensures that \(\sum_{b=1}^T Q_{bs}=Z_s^\top\) and \(\Lambda=\Lambda^\top\). Thus, the precision block update reads
\begin{equation}
 \widetilde{Q}_{ts}-Q_{ts}=\widetilde{Z}_t\Delta Z_s^\top.
 \label{eq:Q-update-row-sums}
\end{equation}
To eliminate \(\widetilde{Z}_t\) in favor of \(Z_t\), fix \(t\) and sum \eqref{eq:Q-update-row-sums} over \(s=1,\ldots,T\). For instance, with \(T=3\) and \(t=2\), summing across blocks \((2,1),(2,2),(2,3)\) yields the difference of the second block row sums on the left while factoring out \(\widetilde{Z}_2\Delta\) on the right. In general,
\begin{align*}
 \widetilde{Z}_t-Z_t
 &=\sum_{s=1}^T(\widetilde{Q}_{ts}-Q_{ts})
   =\widetilde{Z}_t\Delta\sum_{s=1}^T Z_s^\top
   =\widetilde{Z}_t\Delta\Lambda,\\
 \widetilde{Z}_t(I_p-\Delta\Lambda)&=Z_t,\qquad t=1,\ldots,T.
\end{align*}
The matrix \(I_p-\Delta\Lambda\) is necessarily nonsingular. Indeed, if \((I_p-\Delta\Lambda)v=0\) for some \(v\in\R^p\), right-multiplying the preceding relation by \(v\) implies \(Z_t v=0\) for every \(t=1,\ldots,T\). Stacking these equations gives
\begin{equation*}
 Q(\one_T\otimes v)
 =\begin{pmatrix} Z_1v\\ \vdots\\ Z_Tv\end{pmatrix}
 =0.
\end{equation*}
Because \(Q\succ 0\) is invertible, \(\one_T\otimes v=0\), which forces \(v=0\).

Right-multiplying by \((I_p-\Delta\Lambda)^{-1}\) yields \(\widetilde{Z}_t=Z_t(I_p-\Delta\Lambda)^{-1}\). Substituting this back into \eqref{eq:Q-update-row-sums} produces the exact algebraic update
\begin{align}
 \widetilde{Q}_{ts}-Q_{ts}&=Z_t H Z_s^\top,\label{eq:Q-update}\\
 H&=(I_p-\Delta\Lambda)^{-1}\Delta.\label{eq:H}
\end{align}
For every nonadjacent pair \(t\ge s+2\), both \(Q_{ts}\) and \(\widetilde{Q}_{ts}\) must vanish, yielding the exact linear restrictions
\begin{equation}
 Z_t H Z_s^\top=0_{p\times p},\qquad t\ge s+2.
 \label{eq:zero-restrictions}
\end{equation}
Conversely, satisfying \eqref{eq:zero-restrictions} preserves all zero blocks below the first subdiagonal, and symmetry of \(\widetilde{Q}\) preserves those above the first superdiagonal. Provided \(\widetilde{\Omega}\succ 0\) and \(\widetilde{K}\succ 0\), the inverse parameter map then recovers a valid within-person parameterization. Crucially, \(Z_t\) and \(Z_s\) depend solely on the baseline parameter, rendering \eqref{eq:zero-restrictions} a system of homogeneous linear equations in \(H\).

Note that \(H\) is symmetric: because \((I_p-\Delta\Lambda)\Delta=\Delta(I_p-\Lambda\Delta)\), we have
\begin{equation*}
 H=(I_p-\Delta\Lambda)^{-1}\Delta
 =\Delta(I_p-\Lambda\Delta)^{-1}=H^\top.
\end{equation*}
To invert the transformation from \(H\) back to \(\Delta\), left-multiply \eqref{eq:H} by \(I_p-\Delta\Lambda\) to obtain \(H=\Delta+\Delta\Lambda H=\Delta(I_p+\Lambda H)\). Consequently,
\begin{equation*}
 (I_p-\Lambda\Delta)(I_p+\Lambda H)
 =I_p+\Lambda(H-\Delta-\Delta\Lambda H)
 =I_p,
\end{equation*}
so \(I_p+\Lambda H\) is invertible and
\begin{equation}
 \Delta=H(I_p+\Lambda H)^{-1}=(I_p+H\Lambda)^{-1}H.
 \label{eq:Delta}
\end{equation}
The two expressions are transposes of one another for symmetric \(H\), guaranteeing that \(\Delta\) is symmetric. Furthermore, because \(\det(I_p+\Lambda H)\) is continuous in \(H\) and equals \(1\) at \(H=0\), the inverse in \eqref{eq:Delta} exists for all sufficiently small symmetric solutions \(H\) of \eqref{eq:zero-restrictions}.

\subsection{Rank Criterion and Number of Waves}

Applying the vectorization identity \eqref{eq:vec-product} and the duplication relation \(\vecop(H)=\Dp\vech(H)\) to \eqref{eq:zero-restrictions} expresses each nonadjacent zero-block restriction in terms of the distinct entries of \(H\):
\begin{equation}
 \vecop(Z_tHZ_s^\top)
 =(Z_s\otimes Z_t)\vecop(H)
 =(Z_s\otimes Z_t)\Dp\vech(H)=0_{p^2}.
 \label{eq:vec-restriction}
\end{equation}
For example, when \(p=2\), the duplication matrix repeats the off-diagonal element \(h_{21}\):
\begin{equation*}
 \vecop(H)
 =\begin{pmatrix}h_{11}\\h_{21}\\h_{21}\\h_{22}\end{pmatrix}
 =\underbrace{\begin{pmatrix}
  1&0&0\\
  0&1&0\\
  0&1&0\\
  0&0&1
 \end{pmatrix}}_{\mathsf{D}_2}
 \underbrace{\begin{pmatrix}h_{11}\\h_{21}\\h_{22}\end{pmatrix}}_{\vech(H)},
\end{equation*}
so each zero block imposes \(p^2=4\) scalar equations on \(s_p=3\) free parameters.

All nonadjacent pairs \((t,s)\) constrain the same vector \(\vech(H)\). Stack the constraint matrices \((Z_s\otimes Z_t)\Dp\) across all \((T-1)(T-2)/2\) pairs satisfying \(t\ge s+2\) to form the identification Jacobian matrix \(\mathcal{R}_{p,T}\in\R^{\frac{1}{2}(T-1)(T-2)p^2\times s_p}\). (For instance, when \(T=4\), the stacked pairs are \((3,1)\), \((4,1)\), and \((4,2)\).) Thus, the homogeneous system \(\mathcal{R}_{p,T}\vech(H)=0\) admits only the trivial solution \(H=0\) if and only if \(\mathcal{R}_{p,T}\) has full column rank.

\Needspace{7\baselineskip}
\begin{theorem}[Global Identification of the RI-CLPM]
\label{thm:identification}
In the RI-CLPM \eqref{eq:observation}--\eqref{eq:inputs} with transition matrices and innovation covariances free to vary across waves, the covariance parameter vector \(\theta\) is globally identified from \(\Sigma(\theta)\) if and only if
\begin{equation}
 \rank(\mathcal{R}_{p,T})=s_p.
 \label{eq:rank}
\end{equation}
Whenever \(\rank(\mathcal{R}_{p,T})<s_p\), there exists a continuum of admissible parameters arbitrarily close to \(\theta\) that induce the exact same population covariance \(\Sigma(\theta)\).
\end{theorem}

\begin{proof}
Suppose \(\rank(\mathcal{R}_{p,T})=s_p\). For any admissible covariance-equivalent decomposition \((\widetilde{K},\widetilde{\Omega})\), \eqref{eq:Q-update} and \eqref{eq:vec-restriction} force \(\mathcal{R}_{p,T}\vech(H)=0\), hence \(H=0\). Equation \eqref{eq:Delta} then yields \(\Delta=0\), so \(\widetilde{\Omega}=\Omega\) and \(\widetilde{K}=K\). Uniqueness of the within-person inverse parameter map ensures equality of all remaining parameters in \(\theta\).

Conversely, suppose \(\rank(\mathcal{R}_{p,T})<s_p\), and choose a nonzero symmetric matrix \(H_0\) satisfying \eqref{eq:zero-restrictions}. By linearity, \(H=aH_0\) solves \eqref{eq:zero-restrictions} for every scalar \(a\in\R\). For sufficiently small \(\lvert a\rvert\), \eqref{eq:Delta} defines a smooth curve of symmetric perturbations
\begin{equation*}
 \Delta(a)=aH_0(I_p+a\Lambda H_0)^{-1}\longrightarrow 0
 \qquad\text{as }a\to 0.
\end{equation*}
At \(a=0\), \(\widetilde{\Omega}(0)=\Omega\succ 0\) and \(\widetilde{K}(0)=K\succ 0\) have strictly positive minimum eigenvalues. By continuity of eigenvalues, \(\widetilde{\Omega}(a)\succ 0\) and \(\widetilde{K}(a)\succ 0\) for all sufficiently small \(\lvert a\rvert\).

For \(H=aH_0\), the construction of \(\Delta(a)\) satisfies \((I_p-\Delta(a)\Lambda)H=\Delta(a)\). Positive definiteness of \(\widetilde{K}(a)\) guarantees invertibility of \(I_p-\Delta(a)\Lambda\), so \eqref{eq:H} recovers \(H=aH_0\). Thus, \eqref{eq:zero-restrictions} preserves the block-tridiagonal structure of \(\widetilde{Q}(a)\), while \eqref{eq:K-perturb} preserves \(\Sigma\). Every \(a\ne 0\) yields \(\Delta(a)\ne 0\), and distinct values of \(a\) yield distinct perturbations because the mapping \(H\mapsto\Delta\) is invertible. Finally, the recovered within-person parameters converge continuously to their baseline values as \(a\to 0\) via \eqref{eq:within-recovery-from-K}.
\end{proof}

Differentiating at the original parameter values gives the local version of the identification condition. With \(\Sigma\) fixed in \eqref{eq:Sigma}, we have \(\dd K=-(\one_T\one_T^\top)\otimes\dd\Omega\). Differentiating \(QK=I_d\) gives \(\dd Q=-Q(\dd K)Q\). Its \((t,s)\) block, using \eqref{eq:Z}, is
\begin{equation}
 \dd Q=-Q(\dd K)Q,\qquad
 \dd Q_{ts}=Z_t(\dd\Omega)Z_s^\top.
 \label{eq:Q-differential}
\end{equation}
The same linear operator therefore governs covariance-preserving perturbations \(\dd\Omega\) and the global reparameterization \(H\) in \eqref{eq:Q-update}.

Using \eqref{eq:Q-first}--\eqref{eq:Q-zeros}, the block row sums \(Z_t\) take the explicit forms
\begin{align}
 Z_1&=\Gamma^{-1}-F_2^\top\Psi_2^{-1}(I_p-F_2),\label{eq:Z-first}\\
 Z_t&=\Psi_t^{-1}(I_p-F_t)
       -F_{t+1}^\top\Psi_{t+1}^{-1}(I_p-F_{t+1}),
       &&2\le t<T,\label{eq:Z-middle}\\
 Z_T&=\Psi_T^{-1}(I_p-F_T).\label{eq:Z-last}
\end{align}
Under the positive definiteness assumptions in \eqref{eq:inputs}, the rank in \eqref{eq:rank} is determined by the within-person parameters through \(Q\) alone. It does not depend on \(\Omega\).

Specializing Theorem~\ref{thm:identification} to small \(T\) yields immediate structural insights. When \(T=2\), there are no nonadjacent wave pairs, so \(\rank(\mathcal{R}_{p,2})=0\) and the unrestricted RI-CLPM is never identified. When \(T=3\), the sole constraint is \(Z_3 H Z_1^\top=0\). If both \(Z_1\) and \(Z_3\) are invertible, pre-multiplying by \(Z_3^{-1}\) and post-multiplying by \((Z_1^\top)^{-1}\) forces \(H=0\). Conversely, if either \(Z_1\) or \(Z_3\) is singular, selecting a nonzero vector \(v\) in its null space (or that of \(Z_3\) if \(Z_3v=0\), else \(Z_1v=0\)) yields the nonzero symmetric rank-one solution \(H=vv^\top\), since \(Z_3(vv^\top)Z_1^\top=(Z_3v)(Z_1v)^\top=0\). Hence, for \(T=3\), global identification holds if and only if \(\det(Z_1)\ne 0\) and \(\det(Z_3)\ne 0\).

For example, setting \(F_2=I_p\) yields \(Z_1=\Gamma^{-1}\succ 0\). Provided \(I_p-F_3\) is nonsingular, \(Z_3=\Psi_3^{-1}(I_p-F_3)\) is also nonsingular, so three waves suffice for global identification even when one transition matrix equals the identity. Section~\ref{sec:comparison} revisits this exact configuration to link identification to slope equivalence across models.

\subsection{Time-Invariant Transitions}
\label{sec:ident-common}

Now impose time-invariant transitions \(F_t=F\) for all \(t=2,\ldots,T\), with \(T\geq 3\). Within this restricted model, the covariance parameter vector is globally identified if and only if \(\det(I_p-F)\ne 0\).

Under \(F_t=F\), the initial within-person state \(w_{i1}\) propagates to waves \(2\) and \(3\) as \(Fw_{i1}\) and \(F^2w_{i1}\), respectively. Because innovations are orthogonal to \(w_{i1}\), the first-column covariance blocks of \(\Sigma\) satisfy
\begin{equation*}
 \Sigma_{11}=\Omega+\Gamma,\qquad
 \Sigma_{21}=\Omega+F\Gamma,\qquad
 \Sigma_{31}=\Omega+F^2\Gamma.
\end{equation*}
Taking first differences eliminates the common random intercept covariance \(\Omega\):
\begin{align}
 D_1&=\Sigma_{21}-\Sigma_{11}=(F-I_p)\Gamma,\label{eq:common-D1}\\
 D_2&=\Sigma_{31}-\Sigma_{21}=F(F-I_p)\Gamma=FD_1.
 \label{eq:common-D2}
\end{align}
Both \(D_1\) and \(D_2\) are directly observed from \(\Sigma\). Since \(\Gamma\succ 0\), \(D_1=(F-I_p)\Gamma\) is invertible if and only if \(\det(I_p-F)\ne 0\). When this holds, right-multiplying \eqref{eq:common-D2} by \(D_1^{-1}\) uniquely recovers \(F=D_2D_1^{-1}\).

Once \(F\) is known, subtracting \(F\Sigma_{11}=F\Omega+F\Gamma\) from \(\Sigma_{21}=\Omega+F\Gamma\) cancels \(F\Gamma\) to yield \(\Sigma_{21}-F\Sigma_{11}=(I_p-F)\Omega\), whence
\begin{equation}
 \Omega=(I_p-F)^{-1}(\Sigma_{21}-F\Sigma_{11}).
 \label{eq:common-recovery}
\end{equation}
Subtracting \(\Omega\) from the remaining blocks of \(\Sigma\) determines \(K\), after which \(\Gamma=K_{11}\) and \eqref{eq:within-recovery-from-K} uniquely recover all \(\Psi_t\). Thus, \(\det(I_p-F)\ne 0\) is sufficient for global identification regardless of whether \(\Psi_t\) varies across waves.

To establish necessity, suppose \(\det(I_p-F)=0\), so there exists a nonzero vector \(v\in\R^p\) satisfying \(Fv=v\) (and hence \(F^jv=v\) for all \(j\ge 0\)). Perturb \(\widetilde{\Omega}=\Omega+a vv^\top\) and \(\widetilde{\Gamma}=\Gamma-a vv^\top\) while holding \(F\) and all \(\Psi_t\) fixed. For sufficiently small \(a>0\), both \(\widetilde{\Omega}\succ 0\) and \(\widetilde{\Gamma}\succ 0\). Because the innovation contributions to \(K_{ts}\) are unaltered, every within-person block shifts by
\begin{align*}
 \widetilde{K}_{ts}-K_{ts}
 &=F^{t-1}(-a vv^\top)(F^{s-1})^\top\\
 &=-a(F^{t-1}v)(F^{s-1}v)^\top=-a vv^\top,
\end{align*}
which is exactly offset by adding \(a vv^\top\) to \(\Omega\) in \(\Sigma_{ts}=\Omega+K_{ts}\). Substantively, along any unit-eigenvalue direction of \(F\), initial within-person variance is indistinguishable from time-invariant between-person variance, proving nonidentification.

\textcite{Hamaker2015RICLPM} give three waves as the minimum for the RI-CLPM. With a common transition matrix, stationarity imposes \(\rho(F)<1\) and therefore makes \(I_p-F\) nonsingular. The argument above proves global identification of the covariance parameters in the stationary RI-CLPM for \(T\geq 3\).
% End chapters/02_identification.tex
% Begin chapters/03_estimation.tex
\section{Estimation}
\label{sec:estimation}

Estimation methods for structural equation models include generalized
and weighted least squares \parencite{Cheung2013REML}, Bayesian
estimation \parencite{Ulitzsch2023SmallSampleSEM}, and model-implied
instrumental variables for cross-lagged models \parencite{Fang2025MIIV}.
Maximum likelihood (ML) is the conventional choice for SEMs with
continuous variables \parencite{Cheung2013REML,Ulitzsch2023SmallSampleSEM}.
We use Gaussian ML for the RI-CLPM specified in Section~\ref{sec:model}
and derive its score equations for complete trajectories.

We assume throughout this section that the profile log-likelihood has a maximizer in the interior of the parameter space. For the general theory of maximum likelihood and estimating equations, see \textcite{VanDerVaart1998}.

Three parts of the covariance representation determine the scores. The covariance \(\Sigma\) depends on the transitions through \(\mathcal{L}\), on the initial and innovation covariances through \(\mathcal{D}\), and on the random intercept covariance through the Kronecker term in \eqref{eq:Sigma}. We differentiate each part to obtain a closed-form score for its parameter block.

\subsection{Gaussian Log-Likelihood and Profiling of Wave Means}

Let \(\bar{y}=N^{-1}\sum_{i=1}^N y_i\) denote the sample mean trajectory, \(S=N^{-1}\sum_{i=1}^N(y_i-\bar{y})(y_i-\bar{y})^\top\) the maximum-likelihood sample covariance matrix, and \(S_\mu=N^{-1}\sum_{i=1}^N(y_i-\mu)(y_i-\mu)^\top\) the second-moment matrix centered at \(\mu\). Expanding \(y_i-\mu=(y_i-\bar{y})+(\bar{y}-\mu)\) eliminates the cross terms and gives
\begin{equation}
 S_\mu=S+(\bar{y}-\mu)(\bar{y}-\mu)^\top.
 \label{eq:sample-covariance}
\end{equation}
For \(N\) independent trajectories, each with density
\begin{equation}
 f(y_i;\mu,\theta)=(2\pi)^{-d/2}(\det\Sigma)^{-1/2}
 \exp\!\left[-\tfrac{1}{2}(y_i-\mu)^\top P(y_i-\mu)\right],
 \label{eq:trajectory-density}
\end{equation}
the joint log-likelihood function is
\begin{equation}
 \ell(\mu,\theta)
 =c_N-\frac{N}{2}\left[\log\det\Sigma+\tr(P S_\mu)\right],
 \qquad c_N=-\frac{Nd}{2}\log(2\pi),
 \label{eq:likelihood}
\end{equation}
where we used \((y_i-\mu)^\top P(y_i-\mu)=\tr[P(y_i-\mu)(y_i-\mu)^\top]\).

For any admissible \(\theta\), \(P=\Sigma^{-1}\succ 0\), so \eqref{eq:likelihood} depends on \(\mu\) solely through the negative definite quadratic form \(-\frac{N}{2}(\bar{y}-\mu)^\top P(\bar{y}-\mu)\). Its differential is \(N(\bar{y}-\mu)^\top P\,\dd\mu\), uniquely maximizing \(\ell(\mu,\theta)\) at \(\widehat{\mu}=\bar{y}\). Substituting \(\widehat{\mu}=\bar{y}\) into \eqref{eq:likelihood} yields the profile log-likelihood~for~\(\theta\):
\begin{equation}
 \ell_p(\theta)=c_N-\frac{N}{2}\left[\log\det\Sigma+\tr(P S)\right].
 \label{eq:profile}
\end{equation}
We hold the sample covariance matrix \(S\) fixed in all subsequent derivatives.

\subsection{Total Differential of the Profile Log-Likelihood}

Differentiating \(P\Sigma=I_d\) gives \(\dd P=-P(\dd\Sigma)P\), while Jacobi's formula gives \(\dd\log\det\Sigma=\tr(P\,\dd\Sigma)\). Applying both identities to \eqref{eq:profile} and invoking cyclic invariance of the trace yields
\begin{align}
 \dd\ell_p
 &=-\frac{N}{2}\left\{\tr(P\,\dd\Sigma)+\tr[(\dd P)S]\right\}\nonumber\\
 &=\frac{N}{2}\left\{\tr(PSP\,\dd\Sigma)-\tr(P\,\dd\Sigma)\right\}\nonumber\\
 &=\frac{N}{2}\tr(\mathcal{G}\,\dd\Sigma),
 \qquad \mathcal{G}=P(S-\Sigma)P,
 \label{eq:likelihood-differential}
\end{align}
where the final equality follows from \(P\Sigma P=P\). The differential \(\dd\ell_p\) is linear in the parameter differentials. To obtain the score for one block, set the other differentials to zero.

Differentiating the additive decomposition \eqref{eq:Sigma} gives
\begin{equation}
 \dd\Sigma=\dd K+(\one_T\one_T^\top)\otimes\dd\Omega.
 \label{eq:Sigma-differential}
\end{equation}
To differentiate \(K=\mathcal{L}^{-1}\mathcal{D}(\mathcal{L}^{-1})^\top\), set \(C=\mathcal{L}^{-1}\) temporarily so that \(\dd C=-C(\dd\mathcal{L})C\). Expanding \(\dd K=(\dd C)\mathcal{D}C^\top+C(\dd\mathcal{D})C^\top+C\mathcal{D}(\dd C)^\top\) yields
\begin{equation}
 \begin{aligned}
 \dd K
 &=\mathcal{L}^{-1}(\dd\mathcal{D})(\mathcal{L}^{-1})^\top\\
 &\quad-\mathcal{L}^{-1}(\dd\mathcal{L})K
 -K(\dd\mathcal{L})^\top(\mathcal{L}^{-1})^\top.
 \end{aligned}
 \label{eq:K-differential}
\end{equation}
The first term accounts for changes in \(\Gamma\) and \(\Psi_t\). The remaining two terms account for changes in \(F_t\).

\subsection{Score for the Random Intercept Covariance}

Setting \(\dd K=0\) in \eqref{eq:Sigma-differential} gives \(\dd\Sigma=(\one_T\one_T^\top)\otimes\dd\Omega\), so every \(p\times p\) block of \(\dd\Sigma\) equals \(\dd\Omega\). Consequently, the \(r\)th diagonal block of \(\mathcal{G}\,\dd\Sigma\) is \(\sum_{s=1}^T\mathcal{G}_{rs}\,\dd\Omega\). Summing the traces of all diagonal blocks in \eqref{eq:likelihood-differential} gives
\begin{equation}
 \dd\ell_p=\frac{N}{2}\tr(A_\Omega\,\dd\Omega),
 \qquad A_\Omega=\sum_{r=1}^T\sum_{s=1}^T\mathcal{G}_{rs}.
 \label{eq:omega-trace}
\end{equation}
Since \(\mathcal{G}\) is symmetric (\(\mathcal{G}_{sr}=\mathcal{G}_{rs}^\top\)), \(A_\Omega\) is symmetric. Expressing \(\dd\Omega\) in terms of the free parameter vector \(\omega=\vech(\Omega)\) via \eqref{eq:trace-duplication} yields
\begin{equation}
 \dd\ell_p=\frac{N}{2}\vecop(A_\Omega)^\top\Dp\,\dd\omega,
 \qquad
 \boxed{U_\omega=\frac{N}{2}\Dp^\top\vecop(A_\Omega).}
 \label{eq:score-omega}
\end{equation}
Here \(\Dp^\top\) accounts for symmetry by summing the partial derivatives corresponding to duplicate off-diagonal entries (e.g., entries \((1,2)\) and \((2,1)\) for \(\omega_{12}\)).

\subsection{Scores for the Initial and Innovation Covariances}

Next, set \(\dd\Omega=0\) and \(\dd\mathcal{L}=0\). Combining \eqref{eq:likelihood-differential} with the first term of \eqref{eq:K-differential} and cycling the trace gives
\begin{equation}
 \dd\ell_p=\frac{N}{2}\tr(\mathcal{A}\,\dd\mathcal{D}),
 \qquad
 \mathcal{A}=(\mathcal{L}^{-1})^\top\mathcal{G}\mathcal{L}^{-1},
 \label{eq:A-score}
\end{equation}
where \(\mathcal{A}\) is symmetric. Varying \(\Gamma\) alone places \(\dd\Gamma\) in block \((1,1)\) of \(\dd\mathcal{D}\) with zeros elsewhere, reducing the trace to \(\tr(\mathcal{A}_{11}\,\dd\Gamma)\). Similarly, varying \(\Psi_t\) alone places \(\dd\Psi_t\) in block \((t,t)\) of \(\dd\mathcal{D}\), yielding \(\tr(\mathcal{A}_{tt}\,\dd\Psi_t)\). Applying \eqref{eq:trace-duplication} gives the block scores with respect to \(\vech(\Gamma)\) and \(\vech(\Psi_t)\):
\begin{align}
 &\boxed{U_\Gamma=\frac{N}{2}\Dp^\top\vecop(\mathcal{A}_{11})},\label{eq:score-gamma}\\
 &\boxed{U_{\Psi_t}=\frac{N}{2}\Dp^\top\vecop(\mathcal{A}_{tt})},
 \qquad t=2,\ldots,T.
 \label{eq:score-psi}
\end{align}

\subsection{Scores for the Transition Matrices}

Finally, set \(\dd\Omega=0\) and \(\dd\mathcal{D}=0\). Because \(\mathcal{G}\) and \(K\) are symmetric, the two transition terms in \eqref{eq:K-differential} produce identical traces:
\begin{equation}
 \begin{aligned}
 \tr\!\left[\mathcal{G} K(\dd\mathcal{L})^\top(\mathcal{L}^{-1})^\top\right]
 &=\tr\!\left[K\mathcal{G}\mathcal{L}^{-1}\dd\mathcal{L}\right]\\
 &=\tr\!\left[\mathcal{G}\mathcal{L}^{-1}(\dd\mathcal{L})K\right].
 \end{aligned}
 \label{eq:F-two-traces}
\end{equation}
Combining both contributions in \eqref{eq:likelihood-differential} yields
\begin{equation}
 \dd\ell_p=-N\tr(\mathcal{C}^\top\dd\mathcal{L}),
 \qquad \mathcal{C}=(\mathcal{L}^{-1})^\top\mathcal{G} K.
 \label{eq:transition-trace}
\end{equation}
When only \(F_t\) varies, the sole nonzero block of \(\dd\mathcal{L}\) is \(-\dd F_t\) in position \((t,t-1)\), canceling the leading minus sign in \eqref{eq:transition-trace}:
\begin{gather}
 \dd\ell_p
 =N\tr\!\left(\mathcal{C}_{t,t-1}^\top\dd F_t\right)
 =N\vecop(\mathcal{C}_{t,t-1})^\top\dd\vecop(F_t),\nonumber\\
 \boxed{U_{F_t}=N\vecop(\mathcal{C}_{t,t-1})},
 \qquad t=2,\ldots,T.
 \label{eq:score-F}
\end{gather}
For example, if \(F_t=\bigl(\begin{smallmatrix}a&b\\c&d\end{smallmatrix}\bigr)\) and only \(b\) varies, \(\dd F_t=\bigl(\begin{smallmatrix}0&1\\0&0\end{smallmatrix}\bigr)\dd b\) isolates the \((1,2)\) entry \(N(\mathcal{C}_{t,t-1})_{12}\); the full score vector \(U_{F_t}\) stacks all \(p^2\) entries simultaneously in columnwise order.

\subsection{First-Order Conditions for Maximum Likelihood}

Let \(\widehat{\theta}\) collect the estimated covariance and transition parameters, and let hats on \(A_\Omega\), \(\mathcal{A}\), and \(\mathcal{C}\) denote evaluation at \(\widehat{\theta}\) with \(S\) fixed. With wave-specific parameters free, the profiled mean estimate and the zero-score conditions in \eqref{eq:score-omega}, \eqref{eq:score-gamma}, \eqref{eq:score-psi}, and \eqref{eq:score-F} give
\begin{equation*}
 \boxed{\begin{aligned}
  \widehat{\mu}&=\bar{y},\\
  \widehat{A}_{\Omega}&=0_{p\times p},\\
  \widehat{\mathcal{A}}_{11}&=0_{p\times p},\\
  \widehat{\mathcal{A}}_{tt}&=0_{p\times p}, &&t=2,\ldots,T,\\
  \widehat{\mathcal{C}}_{t,t-1}&=0_{p\times p}, &&t=2,\ldots,T.
 \end{aligned}}
\end{equation*}
For a symmetric matrix \(M\), \(\Dp^\top\vecop(M)=\Dp^\top\Dp\vech(M)\). Since \(\Dp^\top\Dp\) is nonsingular, a zero covariance score is equivalent to a zero matrix in the corresponding condition. The transition condition follows from the one-to-one vectorization of \(\mathcal{C}_{t,t-1}\).

If a parameter block is constrained to be equal across waves, the chain rule sums its wave-specific score contributions. For a common transition matrix \(F\) or a common innovation covariance \(\Psi\), the respective equations are
\begin{equation*}
 \sum_{t=2}^{T}\widehat{\mathcal{C}}_{t,t-1}=0_{p\times p},
 \qquad
 \sum_{t=2}^{T}\widehat{\mathcal{A}}_{tt}=0_{p\times p}.
\end{equation*}
% End chapters/03_estimation.tex
% Begin chapters/04_model_comparison.tex
\section{Model Comparison at the Population Level}
\label{sec:comparison}

The conventional CLPM relates the observed variables at wave \(t\) to those at wave \(t-1\) through the equation
\begin{equation*}
 y_{it}=a_t+B_t y_{i,t-1}+r_{it},\qquad t=2,\ldots,T.
\end{equation*}
In this regression, \(a_t\in\R^p\) is the intercept, \(r_{it}\) is the residual, and \(B_t\in\R^{p\times p}\) is the slope matrix. The diagonal entries of \(B_t\) are autoregressive slopes, and its off-diagonal entries are cross-lagged slopes. Our comparison relates the population value of \(B_t\), written \(\mathcal{B}_t\), to the transition matrix \(F_t\) for successive within-person deviations \parencite{Hamaker2015RICLPM,Hamaker2026WithinBetween}.

Take the RI-CLPM in \eqref{eq:observation}--\eqref{eq:inputs} as the true data-generating process, with fixed parameter values that determine the population means and covariances of the observations. Regressing \(y_{it}\) on \(y_{i,t-1}\) in that population yields \(\mathcal{B}_t=\Sigma_{t,t-1}\Sigma_{t-1,t-1}^{-1}\), with both covariance blocks supplied by \eqref{eq:adjacent-blocks}. The intercept follows from the slope and wave means. The initial covariance comes from the first wave, and the innovation covariance comes from the regression residuals.

\subsection{Wave-Specific Population CLPM Parameters}

Write the regression in terms of a stacked parameter vector, a design matrix, and a residual vector.
\begin{equation}
 \begin{gathered}
 \vartheta_t=\begin{pmatrix}a_t\\ \vecop(B_t)\end{pmatrix},
 \qquad
 \mathsf{X}_{it}=\begin{bmatrix} I_p & y_{i,t-1}^\top\otimes I_p\end{bmatrix},\\
 r_{it}=y_{it}-\mathsf{X}_{it}\vartheta_t.
 \end{gathered}
 \label{eq:clpm-design}
\end{equation}
Here \(\mathsf{X}_{it}\in\R^{p\times(p+p^2)}\), and \eqref{eq:vec-product} gives \(\mathsf{X}_{it}\vartheta_t=a_t+B_t y_{i,t-1}\). For instance, when \(p=2\) and \(y_{i,t-1}=(x_1,x_2)^\top\), the slope component expands as
\begin{equation}
 (y_{i,t-1}^\top\otimes I_2)\vecop(B_t)
 =
 \begin{pmatrix}x_1&0&x_2&0\\0&x_1&0&x_2\end{pmatrix}
 \begin{pmatrix}b_{11}\\ b_{21}\\ b_{12}\\ b_{22}\end{pmatrix},
 \label{eq:clpm-design-example}
\end{equation}
where each row corresponds to an outcome variable and each column corresponds to a vectorized slope coefficient.

The population CLPM intercept and slope minimize the expected squared prediction error under the RI-CLPM
\begin{equation}
 J_t(\vartheta_t)=\E_{\mathrm{RI}}[r_{it}^\top r_{it}],
 \label{eq:ls-objective}
\end{equation}
where \(\E_{\mathrm{RI}}[\cdot]\) denotes expectation under the fixed RI-CLPM distribution. Differentiating with respect to \(\vartheta_t\) using \(\dd r_{it}=-\mathsf{X}_{it}\,\dd\vartheta_t\) gives the first-order optimality condition
\begin{align}
 \dd J_t&=-2\,\E_{\mathrm{RI}}[r_{it}^\top\mathsf{X}_{it}]\,\dd\vartheta_t,\label{eq:ls-diff}\\
 \E_{\mathrm{RI}}[\mathsf{X}_{it}^\top r_{it}^\star]
 &=\begin{pmatrix}
 \E_{\mathrm{RI}}[r_{it}^\star]\\
 \vecop\{\E_{\mathrm{RI}}[r_{it}^\star y_{i,t-1}^\top]\}
 \end{pmatrix}
 =0_{(p+p^2)\times 1},
 \label{eq:ls-stacked-normal}
\end{align}
where the lower block uses \((y_{i,t-1}\otimes I_p)r_{it}^\star=\vecop(r_{it}^\star y_{i,t-1}^\top)\).

Denote the population CLPM slope by \(\mathcal{B}_t\). The upper block of \eqref{eq:ls-stacked-normal} gives the population CLPM intercept \(a_t^\star=\mu_t-\mathcal{B}_t\mu_{t-1}\). Substituting \(a_t^\star\) into the lower block centers the cross-products around the wave means:
\begin{align}
 \E_{\mathrm{RI}}[r_{it}^\star y_{i,t-1}^\top]
 &=
 \bigl(\E_{\mathrm{RI}}[y_{it}y_{i,t-1}^\top]-\mu_t\mu_{t-1}^\top\bigr)
 \nonumber\\
 &\quad-\mathcal{B}_t\bigl(\E_{\mathrm{RI}}[y_{i,t-1}y_{i,t-1}^\top]
             -\mu_{t-1}\mu_{t-1}^\top\bigr)\nonumber\\
 &=\Sigma_{t,t-1}-\mathcal{B}_t\Sigma_{t-1,t-1}=0_{p\times p}.
 \label{eq:ls-covariance-normal}
\end{align}
Vectorizing this normal equation gives
\begin{align}
 (\Sigma_{t-1,t-1}\otimes I_p)\vecop(\mathcal{B}_t)
 &=\vecop(\Sigma_{t,t-1}),\label{eq:ls-vec-normal}\\
 \vecop(\mathcal{B}_t)
 &=(\Sigma_{t-1,t-1}^{-1}\otimes I_p)\vecop(\Sigma_{t,t-1}).
 \label{eq:ls-vec-solution}
\end{align}
The covariance \(\Sigma_{t-1,t-1}\succ 0\) by \eqref{eq:Sigma}, so \(J_t(\vartheta_t)\) is strictly convex and has a unique global minimizer. To obtain the population slope in matrix form, unvectorize \eqref{eq:ls-vec-solution} and insert the adjacent-wave blocks from \eqref{eq:adjacent-blocks}. This gives
\begin{equation}
 \boxed{\begin{aligned}
 \mathcal{B}_t
 &=\Sigma_{t,t-1}\Sigma_{t-1,t-1}^{-1}\\
 &=(\Omega+F_tK_{t-1,t-1})(\Omega+K_{t-1,t-1})^{-1}.
 \end{aligned}}
 \label{eq:population-B}
\end{equation}

Similarly, the first-wave CLPM covariance is \(\Gamma_{\mathrm{C}}^{\star}=\Sigma_{11}\), and for \(t\ge 2\) the population CLPM innovation covariance equals the residual covariance \(\Psi_{\mathrm{C},t}^{\star}=\Cov(r_{it}^\star,r_{it}^\star)\). Because \(\Cov(r_{it}^\star,y_{i,t-1})=0\) by \eqref{eq:ls-stacked-normal}, taking the covariance of \(r_{it}^\star\) with \(y_{it}=a_t^\star+\mathcal{B}_t y_{i,t-1}+r_{it}^\star\) gives \(\Cov(r_{it}^\star,r_{it}^\star)=\Cov(r_{it}^\star,y_{it})\), or
\begin{equation}
 \Psi_{\mathrm{C},t}^{\star}
 =\Sigma_{tt}-\mathcal{B}_t\Sigma_{t-1,t}
 =\Sigma_{tt}-\Sigma_{t,t-1}\Sigma_{t-1,t-1}^{-1}\Sigma_{t-1,t}.
 \label{eq:clpm-innovation}
\end{equation}

\subsection{Exact Slope Discrepancy Between Models}

To isolate the discrepancy \(\mathcal{B}_t-F_t\), subtract \(F_t\Sigma_{t-1,t-1}\) from \(\mathcal{B}_t\Sigma_{t-1,t-1}=\Sigma_{t,t-1}\) and substitute \eqref{eq:adjacent-blocks}:
\begin{align}
 (\mathcal{B}_t-F_t)\Sigma_{t-1,t-1}
 &=\Sigma_{t,t-1}-F_t\Sigma_{t-1,t-1}\nonumber\\
 &=\Omega+F_tK_{t-1,t-1}-F_t(\Omega+K_{t-1,t-1})\nonumber\\
 &=(I_p-F_t)\Omega.
 \label{eq:difference-before-inverse}
\end{align}
The within-person cross-covariance terms \(F_t K_{t-1,t-1}\) cancel identically. Right-multiplying by \(\Sigma_{t-1,t-1}^{-1}=(\Omega+K_{t-1,t-1})^{-1}\) establishes the exact matrix identity
\begin{equation}
 \boxed{\mathcal{B}_t-F_t=(I_p-F_t)\Omega(\Omega+K_{t-1,t-1})^{-1},}
 \label{eq:population-difference}
\end{equation}
or, equivalently, in vectorized form,
\begin{equation}
 \vecop(\mathcal{B}_t-F_t)
 =(\Sigma_{t-1,t-1}^{-1}\otimes I_p)\vecop((I_p-F_t)\Omega).
 \label{eq:vec-difference}
\end{equation}
Notice how \(K_{t-1,t-1}\) enters the denominator matrix \(\Omega+K_{t-1,t-1}\), accumulating all past innovation covariances through \eqref{eq:K-diag}.

When \(\Omega\succ 0\), \eqref{eq:difference-before-inverse} shows that the entire slope matrix coincides across models (\(\mathcal{B}_t=F_t\)) if and only if \(F_t=I_p\). Individual entries coincide when the corresponding entry of the discrepancy matrix in \eqref{eq:population-difference} is zero. This equality condition links directly back to our identification analysis: for \(T=3\), setting \(F_2=I_p\) (with \(\det(I_p-F_3)\ne 0\)) makes \(\mathcal{B}_2=F_2\) while keeping \(Z_1=\Gamma^{-1}\) and \(Z_3=\Psi_3^{-1}(I_p-F_3)\) nonsingular, so the full RI-CLPM remains globally identified by Theorem~\ref{thm:identification}. Conversely, if \(F_t=I_p\) across all waves, then \(Z_2=\cdots=Z_T=0\), \(\rank(\mathcal{R}_{p,T})=0\), and \(\Omega\) and \(\Gamma\) merge into an unidentifiable sum.

\subsection{Slopes Pooled Across Waves}

Keep the data-generating RI-CLPM fixed and impose a common slope matrix \(\mathcal{B}_c\) and innovation covariance \(\Psi_{\mathrm{C}}^{\star}\) over the \(m=T-1\) CLPM transitions, with separate intercepts \(a_t^c\) for each wave. Minimizing the unweighted sum of interval squared errors \(\sum_{t=2}^T J_t(\vartheta_t)\) under \(B_2=\cdots=B_T=\mathcal{B}_c\) yields \(a_t^c=\mu_t-\mathcal{B}_c\mu_{t-1}\) and combines the lower-block normal equations over \(t=2,\ldots,T\). With \(A=\sum_{t=2}^T\Sigma_{t-1,t-1}\), the combined equation is
\begin{align}
 (A\otimes I_p)\vecop(\mathcal{B}_c)
 &=\sum_{t=2}^T\vecop(\Sigma_{t,t-1})\nonumber\\
 &=m\vecop(\Omega)
    +\sum_{t=2}^T(K_{t-1,t-1}\otimes I_p)\vecop(F_t).
 \label{eq:common-vec}
\end{align}
Unvectorizing \eqref{eq:common-vec} yields the pooled population CLPM slope,
\begin{equation}
 \boxed{\mathcal{B}_c=
 \left(m\Omega+\sum_{t=2}^T F_tK_{t-1,t-1}\right)
 \left(m\Omega+\sum_{t=2}^T K_{t-1,t-1}\right)^{-1}.}
 \label{eq:common-B}
\end{equation}
At this slope and the corresponding wave-specific intercepts, let
\(V_t=\Cov(r_{it}^c,r_{it}^c)\) denote the residual covariance under the
data-generating RI-CLPM. Each residual has mean zero. Its covariance is
positive definite because the adjacent-wave covariance is positive definite
and the coefficient on \(y_{it}\) in the residual is \(I_p\).
For a common innovation precision \(W\succ0\), twice the expected negative
log-likelihood of the Gaussian CLPM transitions, up to an additive constant, is
\[
 -m\log\det W+\tr\!\left(W\sum_{t=2}^T V_t\right).
\]
This objective is strictly convex in \(W\). Its first-order condition is
\(mW^{-1}=\sum_{t=2}^T V_t\), yielding the common innovation covariance
\begin{equation}
 \Psi_{\mathrm{C}}^{\star}
 =\frac{1}{m}\sum_{t=2}^T \Cov(r_{it}^c,r_{it}^c),
 \qquad r_{it}^c=y_{it}-a_t^c-\mathcal{B}_c y_{i,t-1}.
 \label{eq:common-clpm-innovation}
\end{equation}
A fixed positive definite outcome weight left-multiplies the summed normal equation and leaves the solution in \eqref{eq:common-B} unchanged. Weights that vary by interval change the relative contributions of the intervals.

If the data-generating RI-CLPM also obeys time-invariant transitions \(F_t=F\), subtracting \((A\otimes I_p)\vecop(F)\) from \eqref{eq:common-vec} cancels all \(\sum_{t=2}^T(K_{t-1,t-1}\otimes I_p)\vecop(F)\) terms, leaving \mbox{\(m\vecop((I_p-F)\Omega)\)}.

Defining the time-averaged within-person covariance
\[
 \overline{K}=m^{-1}\sum_{t=2}^T K_{t-1,t-1}
\]
so that \(A=m(\Omega+\overline{K})\), we obtain
\begin{equation}
 \boxed{\mathcal{B}_c-F=(I_p-F)\Omega(\Omega+\overline{K})^{-1}.}
 \label{eq:common-difference}
\end{equation}
In the slope difference under time-invariant RI-CLPM transitions, pooling replaces the within-person covariance \(K_{t-1,t-1}\) in \eqref{eq:population-difference} with \(\overline{K}\), the arithmetic average of the within-person covariance blocks across predictor waves. This relation also holds when the within-person process is nonstationary.

\subsection{Bivariate Cross-Lagged Paths: Appearance, Cancellation, Reversal, and Ordering}

Specialize to \(p=2\), labeling the first variable \(X\) and the second \(Y\). Write \(A_t=\Sigma_{t-1,t-1}=(a_{jk})\) and \(C_t=\Sigma_{t,t-1}=(c_{jk})\). The cross-lagged path from \(X_{t-1}\) to \(Y_t\) corresponds to entry \((2,1)\), which occupies the second position in \(\vecop(\cdot)\). Reading off the second row \((0,a_{22},0,-a_{12})/\det A_t\) of \(A_t^{-1}\otimes I_2\) in \eqref{eq:ls-vec-solution} gives
\begin{equation}
 (\mathcal{B}_t)_{21}=
 \frac{a_{22}c_{21}-a_{12}c_{22}}{\det A_t}.
 \label{eq:bivariate-slope}
\end{equation}

Let \(g=(F_t)_{21}\) and \(h=(F_t)_{22}\), and denote the distinct entries of \(K_{t-1,t-1}\) and \(\Omega\) by \((k_x,k_y,k_{xy})\) and \((\omega_x,\omega_y,\omega_{xy})\), respectively. The second and fourth entries of \(\vecop((I_2-F_t)\Omega)\) are \((1-h)\omega_{xy}-g\omega_x\) and \((1-h)\omega_y-g\omega_{xy}\). Applying the same Kronecker row to \eqref{eq:vec-difference} and substituting \(a_{22}=\omega_y+k_y\) and \(a_{12}=\omega_{xy}+k_{xy}\) (after canceling \((1-h)\omega_y\omega_{xy}\)) yields
\begin{align}
 \det(A_t)\bigl[(\mathcal{B}_t)_{21}-g\bigr]
 &=a_{22}\bigl[(1-h)\omega_{xy}-g\omega_x\bigr]\nonumber\\
 &\quad-a_{12}\bigl[(1-h)\omega_y-g\omega_{xy}\bigr]\nonumber\\
 &=(1-h)(\omega_{xy}k_y-\omega_y k_{xy})\nonumber\\
 &\quad-g\{\omega_x(\omega_y+k_y)-\omega_{xy}(\omega_{xy}+k_{xy})\}.
 \label{eq:bivariate-difference}
\end{align}
In particular, when the true within-person cross-lagged effect is zero (\(g=0\)), \eqref{eq:bivariate-difference} reduces to
\begin{equation}
 (\mathcal{B}_t)_{21}=
 \frac{(1-h)(\omega_{xy}k_y-\omega_y k_{xy})}{\det(\Omega+K_{t-1,t-1})}.
 \label{eq:zero-F}
\end{equation}
For \(h\ne 1\), \((\mathcal{B}_t)_{21}\) vanishes if and only if the between-person and within-person regression slopes of \(X\) on \(Y\) coincide (\(\omega_{xy}/\omega_y=k_{xy}/k_y\)). Whenever these ratios differ—for instance, when \(k_{xy}=0\) while \(\omega_{xy}\ne 0\)—a spurious cross-lagged path appears in the CLPM despite \(g=0\). Interchanging \(X\) and \(Y\) yields the symmetric formula for \((\mathcal{B}_t)_{12}\). Moreover, in the proportional-covariance special case \(\Omega=c K_{t-1,t-1}\) (\(c>0\)), \eqref{eq:population-B} simplifies directly to \(\mathcal{B}_t=(c I_p+F_t)/(c+1)\).

A stationary bivariate family illustrates path appearance, exact cancellation, sign reversal, and relative-ordering reversal simultaneously. Consider
\begin{equation}
 F=\begin{pmatrix}1/2&0\\g&1/2\end{pmatrix},\quad
 \Gamma=I_2,\quad
 \Omega=\begin{pmatrix}1&-1/2\\-1/2&1\end{pmatrix},\quad
 \Psi=I_2-FF^\top.
 \label{eq:sign-family}
\end{equation}
Here \(\Psi_{11}=3/4\) and \(\det\Psi=9/16-g^2\), so \(\Psi\succ 0\) whenever \(\lvert g\rvert<3/4\), and \eqref{eq:K-diag} gives \(K_{tt}=I_2\) for all \(t\). Substituting \(a_{22}=2\), \(a_{12}=-1/2\), and \(\det A_t=15/4\) into \eqref{eq:bivariate-difference} and its \(X\leftrightarrow Y\) counterpart yields
\begin{equation}
 (\mathcal{B}_t-F)_{21}=-\frac{7g+1}{15},
 \qquad
 (\mathcal{B}_t)_{21}=\frac{8g-1}{15},
 \qquad
 (\mathcal{B}_t)_{12}=-\frac{1}{15}.
 \label{eq:sign-paths}
\end{equation}
Inspection of \eqref{eq:sign-paths} reveals four distinct qualitative phenomena:
\begin{enumerate}
 \item \emph{Spurious appearance}: At \(g=0\), \(F_{21}=0\) yet \((\mathcal{B}_t)_{21}=-1/15\ne 0\).
 \item \emph{Exact cancellation}: At \(g=1/8\), \(F_{21}=1/8\ne 0\) yet \((\mathcal{B}_t)_{21}=0\).
 \item \emph{Sign reversal}: For \(0<g<1/8\), \(F_{21}>0\) whereas \((\mathcal{B}_t)_{21}<0\).
 \item \emph{Ordering reversal}: At \(g=1/10\), the \(X\to Y\) effect dominates \(Y\to X\) in magnitude within the RI-CLPM (\(\lvert F_{21}\rvert=1/10 > \lvert F_{12}\rvert=0\)), whereas \(Y\to X\) dominates \(X\to Y\) within the CLPM (\(\lvert(\mathcal{B}_t)_{12}\rvert=1/15 > \lvert(\mathcal{B}_t)_{21}\rvert=1/75\)).
\end{enumerate}
Because all within-person marginal variances equal \(1\) and all observed marginal variances equal \(2\) in \eqref{eq:sign-family}, standardizing each model on its own marginal scale preserves these exact magnitude orderings. For a path from variable \(k\) at wave \(t-1\) to variable \(j\) at wave \(t\),
the standardized coefficient equals the unstandardized coefficient times
the predictor standard deviation at \(t-1\) divided by the outcome standard
deviation at \(t\) \parencite{Hamaker2015RICLPM,Schuurman2016CompareCrossLagged}.
The standardized RI-CLPM and population CLPM coefficients are, respectively,
\[
 (F_t)_{jk}\sqrt{\frac{(K_{t-1,t-1})_{kk}}{(K_{tt})_{jj}}}
 \qquad\text{and}\qquad
 (\mathcal{B}_t)_{jk}
 \sqrt{\frac{(\Sigma_{t-1,t-1})_{kk}}{(\Sigma_{tt})_{jj}}}.
\]

\subsection{Connection to Published Bivariate Formulas}

Our matrix identities unify and generalize existing bivariate comparisons in the methodological literature. First, \textcite{Hamaker2015RICLPM} express the standardized CLPM \(X\to Y\) coefficient as \((r_1-r_0 r_y)/(1-r_0^2)\), where \(r_0=\Cor(X_{t-1},Y_{t-1})\), \(r_1=\Cor(Y_t,X_{t-1})\), and \(r_y=\Cor(Y_t,Y_{t-1})\). When the observed variables are standardized to unit variance at each wave, \(a_{11}=a_{22}=1\), \(a_{12}=r_0\), \(c_{21}=r_1\), and \(c_{22}=r_y\); substituting these values into \eqref{eq:bivariate-slope} recovers \((r_1-r_0 r_y)/(1-r_0^2)\) immediately. On an unstandardized scale, substituting \(a_{12}=\omega_{xy}+k_{xy}\), \(c_{21}=\omega_{xy}+g k_x+h k_{xy}\), and \(c_{22}=\omega_y+g k_{xy}+h k_y\) from \eqref{eq:adjacent-blocks} and rescaling by the appropriate standard-deviation ratios reproduces the full covariance expansions in Appendix~B of \textcite{Hamaker2015RICLPM}.

Second, \textcite{RobitzschLudtke2026} examine the unstandardized signed difference \(\delta=(\mathcal{B}_t)_{21}-g\) on a common scale. Their Equation~6 numerator matches the right-hand side of \eqref{eq:bivariate-difference}. Specializing to equal marginal variances across variables (\(k_x=k_y=\sigma^2\), \(k_{xy}=\rho_W\sigma^2\), \(\omega_x=\omega_y=\tau^2\), \(\omega_{xy}=\rho_B\tau^2\)) reduces \eqref{eq:bivariate-difference} to
\begin{align}
 \det(A_t)\delta
 &=(1-h)\sigma^2\tau^2(\rho_B-\rho_W)\nonumber\\
 &\quad-g\tau^2\left\{\sigma^2(1-\rho_B\rho_W)+\tau^2(1-\rho_B^2)\right\},
 \label{eq:equal-variance}
\end{align}
which recovers Equation~7 of \textcite{RobitzschLudtke2026}. Writing the right-hand side of \eqref{eq:equal-variance} as \(C-gM\), where \(C=(1-h)\sigma^2\tau^2(\rho_B-\rho_W)\) and \(M=\tau^2\{\sigma^2(1-\rho_B\rho_W)+\tau^2(1-\rho_B^2)\}>0\) under the positive definiteness conditions in \eqref{eq:inputs}, shows that \((\mathcal{B}_t)_{21}>g\) holds if and only if \(g<C/M\) (with all other parameters fixed). Similarly, comparing absolute magnitudes along the same path gives
\begin{equation}
 \lvert(\mathcal{B}_t)_{21}\rvert>\lvert g\rvert
 \quad\Longleftrightarrow\quad
 (g+\delta)^2-g^2=\delta(2g+\delta)>0.
 \label{eq:abs-comparison}
\end{equation}
Equation~\eqref{eq:abs-comparison} compares the absolute values of the two coefficients for the same lagged path across models. The analysis after \eqref{eq:sign-paths} compares the magnitudes of the two reciprocal lagged paths within each model.

Equations \eqref{eq:population-B} and \eqref{eq:population-difference} apply to all \(p^2\) slope entries for \(p\ge 2\). They retain the wave-specific scope of \textcite{Hamaker2015RICLPM} and extend the time-invariant bivariate comparison of \textcite{RobitzschLudtke2026} to time-varying \(F_t\) and \(\Psi_t\). The pooled slope is given by \eqref{eq:common-B}, and its difference from a common RI-CLPM transition matrix is given by \eqref{eq:common-difference}.
% End chapters/04_model_comparison.tex
% Begin chapters/05_conclusion.tex
\Needspace{7\baselineskip}
\section{Conclusion}
\label{sec:conclusion}

We treat the RI-CLPM as a statistical model and use a common matrix representation to study its identification, likelihood estimation, and relationship to the population CLPM.

For identification, the nonadjacent zero blocks of the block-tridiagonal within-person precision matrix \(Q\) determine which changes in the random intercept covariance \(\Omega\) can be absorbed into the within-person process. The rank condition in Theorem~\ref{thm:identification} characterizes global identification when the transition matrices and innovation covariances can vary across waves. It also explains when three waves suffice and why a unit eigenvalue in a common transition matrix prevents identification of components of the variance decomposition.

For complete Gaussian trajectories, the triangular factorization \(\mathcal{L}\) yields closed-form likelihood scores for each parameter block. A change in \(\Omega\) shifts all observed covariance blocks uniformly. Changing \(\Gamma\) or \(\Psi_t\) affects one diagonal block of \(\mathcal{D}\), and changing \(F_t\) affects both factors of \(K=\mathcal{L}^{-1}\mathcal{D}(\mathcal{L}^{-1})^\top\) symmetrically. This accounts for the different score formulas for \(\Omega\), \(\Gamma\), \(\Psi_t\), and \(F_t\).

For model comparison, fixing the parameters of the data-generating RI-CLPM determines the corresponding population CLPM parameters. Equation~\eqref{eq:population-difference} expresses the slope difference at each transition, and \eqref{eq:common-difference} extends this comparison to pooled slopes when the RI-CLPM transition matrix is time-invariant. The identities recover established bivariate formulas and extend them to multivariate and time-varying settings. They show when paths can appear, cancel, reverse sign, or change their relative ordering across models.

We hope this formal statistical treatment gives methodological researchers a foundation for studying the RI-CLPM in greater depth. The results offer starting points for examining how parameter restrictions affect identification and how alternative estimators behave.
% End chapters/05_conclusion.tex
\printbibliography
\end{document}